\documentclass[pss]{wiley2sp} 
\usepackage{amsmath}
\usepackage{bm}             

\begin{document}

\title{Optical observation of quasiperiodic Heisenberg antiferromagnets in two dimensions}

\author{%
  Takashi Inoue and Shoji Yamamoto\textsuperscript{\Ast}  }

\mail{e-mail
  \textsf{yamamoto@phys.sci.hokudai.ac.jp}}

\institute{%
  Department of Physics, Hokkaido university, Sapporo 060-0810, Japan}


\abstract{\bf%
We calculate magnetic Raman spectra of Heisenberg antiferromagnets 
on the two-dimensional Penrose lattice. 
We follow the Shastry-Shraiman formulation of Raman scattering 
in a strongly correlated Hubbard system and 
obtain the second- and fourth-order effective Raman operators. 
The second-order Raman intensity comes from the $E_{2}$ mode, and 
it is invariant under an arbitrary rotation of polarization vectors. 
The fourth-order Raman intensities consist of $A_{1}$ and $A_{2}$, as well as $E_{2}$, 
modes and therefore yield strong polarization dependence. 
In particular, the $A_{2}$ mode intensity directly detects 
the dynamical spin-chirality fluctuations. 
Employing linearly and circularly polarized lights, 
we can separately extract every irreducible representation from the observations. 
We further discuss effects of magnon-magnon interactions 
on the magnetic Raman scattering. 
Our theory provides a reasonable explanation for 
the two-magnon scattering process. 
}

\maketitle   

\section{Introduction\\}

Since the discovery of quasicrystal \cite{Schehtman}, 
quasiperiodic systems have been of much interest. 
Quasicrystal is characterized by quasiperiodicity, 
which is a long-range order without transrational symmetry 
and a crystallographically forbidden rotational symmetry. 
These expect that the physical properties of quasiperiodic systems are 
quite different from both periodic and amorphous systems. 
Penrose lattice is one of the most popular two-dimensional quasicrystals. 
On this lattice, the tight-binding model for noninteracting electrons has been studied. 
It shows many interesting features such as 
the confined state \cite{Kohmoto,Arai}, 
which is characterized by thermodynamically degenerate states 
with strictly localized and self-similar wave functions, 
and multifractal spectrum \cite{Mace}. 
Recently, quantum critical behavior has been observed in the quasicrystal 
${\rm Au_{51}Al_{34}Yb_{15}}$ \cite{Deguchi}. 
In this compound, the $4f$ electrons of ${\rm Yb}$ are strongly correlated, 
so that investigation of the interplay of the quasiperiodicity and strong correlation 
is a big issue. 
On the quasiperiodic systems, 
strongly correlated electron models have been studied such as 
Hubbard model \cite{Koga}, Ising model for classical spins \cite{Okabe}, and 
Heisenberg model for quantum spins \cite{Szallas,Wessel}. 
In this paper, 
we will study the antiferromagnetic Heisenberg model on the Penrose lattice.

One of the important probes of antiferromagnets is a magnetic Raman scattering. 
It is an inelastic photon scattering mediated by magnetic excitations. 
Loudon and Fleury established the standard framework of the two-magnon Raman scattering 
\cite{Fleury}. 
For instance, it was used to estimate the exchange interaction constant in the 
high-$T_{c}$ superconductor ${\rm La_{2}CuO_{4}}$ 
\cite{Sugai}. 
Theoretically, insulating phase of layered cuprates can be well accounted for 
quasi-two-dimensional Heisenberg antiferromagnets on the square lattice. 
Zero-temperature magnetic Raman spectra are calculated by 
spin-wave theory \cite{Canali,Chubukov,Sandvik}, 
exact diagonalization \cite{Sandvik}, and quantum Monte Carlo method \cite{Sandvik}. 
The magnetic Raman spectrum is also computed in other systems, such as 
the triangle lattice \cite{Perkins} and the Kagome lattice \cite{Lhuillier}. 
Polarization dependence of the magnetic Raman intensity depends on 
the lattice geometry and the symmetry of the ground state. 
It provides useful information of magnetic excitations. 

Microscopic description of the magnetic Raman scattering is given by 
Shastry and Shraiman \cite{ShastryPRL,Shastry}. 
In this formulation, the Loudon-Fleury mechanism is obtained 
in a second-order perturbation theory. 
The higher-order perturbation reads 
beyond the Loudon-Fleury mechanism magnetic Raman scattering, and 
it includes additional magnetic excitations as the spin-chirality terms 
${\bm S}_{i}\cdot({\bm S}_{j}\times{\bm S}_{k})$ and/or 
the ring-exchange terms 
$({\bm S}_{i}\cdot{\bm S}_{j})({\bm S}_{k}\cdot{\bm S}_{l})$ 
\cite{Ko,Michaud}. 
We will present the Raman intensity profile 
within and beyond the Loudon-Fleury mechanism 
on the Penrose lattice Heisenberg antiferromagnets.

\section{Model}

\subsection{Penrose lattice\\}
\begin{figure}[htb]
\includegraphics*[width=\linewidth]{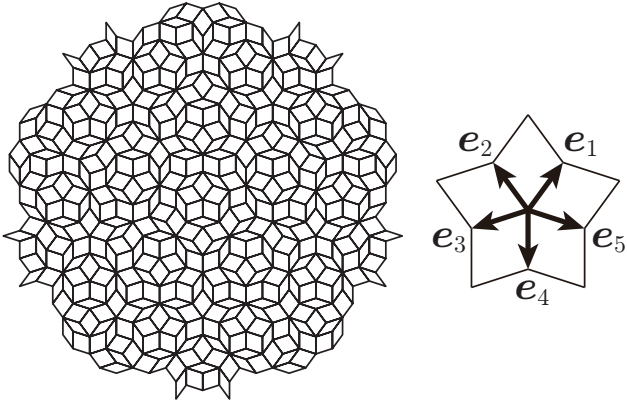}
\caption{
  Central patch of the two-dimensional Penrose lattice with fivefold rotational symmetry 
  and its primitive lattice vectors. 
  ${\bm e}_{1}, \cdots, {\bm e}_{5}$ are projection of 
  the five-dimensional canonical basis vectors, 
  and they satisfy ${\bm e}_{1}+{\bm e}_{2}+{\bm e}_{3}+{\bm e}_{4}+{\bm e}_{5}={\bm 0}$. 
}
\label{penroseimage}
\end{figure}

%
Figure \ref{penroseimage} shows a finite cluster of the Penrose lattice. 
It is composed of two prototiles: 
angle $\pi/5$ (thin) and angle $2\pi/5$ (fat) rhombuses. 
Since the lattice consists of even-number-sided polygons, 
the Penrose lattice is bipartite. 
The two-dimensional Penrose lattice is obtained by 
projection of a five-dimensional hypercubic lattice 
onto an irrational tilted plane \cite{Szallas}, 
and it holds four independent primitive lattice vectors. 
Due to the quasiperiodicity, 
the rank of the Penrose lattice $r=4$ is larger than the lattice dimension $d=2$. 
In this study, we consider open-boundary clusters of the Penrose lattice 
which hold fivefold rotational symmetry.

\subsection{Hamiltonian\\}

We consider the so-called vertex model, 
where spins are located at vertices of the Penrose rhombus tiling. 
We consider the nearest-neighbor antiferromagnetic Heisenberg model: 
\begin{align}
 H=J\sum_{\langle i,j \rangle}{\bm S}_{i}\cdot{\bm S}_{j}
\;\;\;(J>0)
\label{Hamiltonian}
\end{align}
where ${\bm S}_{i}$ is a spin-$1/2$ operator at site $i$, and 
$\langle i,j \rangle$ are pairs of linked vertices of the Penrose lattice.

\subsection{Spin-wave theory\\}

We divide the Penrose lattice into two sublattices A and B 
consisting of $N_{\rm A}$ and $N_{\rm B}$ sites, respectively. 
We introduce bosonic operators by using the Holstein-Primakoff transformation: 
\begin{align}
\nonumber
 S_{i}^{z}&=S-a_{i}^{\dagger}a_{i}
\\ \nonumber
 S_{i}^{+}&=(2S-a_{i}^{\dagger}a_{i})^{\frac{1}{2}}a_{i}
\\
 S_{i}^{-}&=a_{i}^{\dagger}(2S-a_{i}^{\dagger}a_{i})^{\frac{1}{2}}
\end{align}
for $i\in{\rm A}$, and 
\begin{align}
\nonumber
 S_{j}^{z}&=-S+b_{j}^{\dagger}b_{j}
\\ \nonumber
 S_{j}^{+}&=b_{j}^{\dagger}(2S-b_{j}^{\dagger}b_{j})^{\frac{1}{2}}
\\
 S_{j}^{-}&=(2S-b_{j}^{\dagger}b_{j})^{\frac{1}{2}}b_{j}
\end{align}
for $j\in{\rm B}$. 
Expanding the square roots of $1/S$, and keeping terms of $O(S^{0})$, 
spin-wave Hamiltonian is written as
\begin{align}
\nonumber
 H_{\rm SW}
=&J\sum_{\langle i,j \rangle}\biggl[
 -S^{2}
 +S( a_{i}^{\dagger}a_{i}+b_{j}^{\dagger}b_{j}
    +a_{i}b_{j}+a_{i}^{\dagger}b_{j}^{\dagger})
\\
&-\Bigl\{a_{i}^{\dagger}a_{i}b_{j}^{\dagger}b_{j}
  +\frac{1}{4}(a_{i}^{\dagger}a_{i}a_{i}b_{j}
              +a_{i}^{\dagger}b_{j}^{\dagger}b_{j}^{\dagger}b_{j}+{\rm H.c.})
 \Bigr\}\biggr]
\label{SWhamiltonian}
\end{align}
We apply the Wick decomposition for the $O(S^{0})$ terms in 
Eq. \eqref{SWhamiltonian}, 
\begin{align}
\nonumber
 a_{i}^{\dagger}a_{i}b_{j}^{\dagger}b_{j}
\to&
 \langle a_{i}^{\dagger}a_{i} \rangle b_{j}^{\dagger}b_{j}
+\langle b_{j}^{\dagger}b_{j} \rangle a_{i}^{\dagger}a_{i}
-\langle a_{i}^{\dagger}a_{i} \rangle \langle b_{j}^{\dagger}b_{j} \rangle
\\ \nonumber
&+\langle a_{i}^{\dagger}b_{j}^{\dagger} \rangle a_{i}b_{j}
 +\langle a_{i}b_{j} \rangle a_{i}^{\dagger}b_{j}^{\dagger}
 -\langle a_{i}^{\dagger}b_{j}^{\dagger} \rangle \langle a_{i}b_{j} \rangle
\\ \nonumber
 a_{i}^{\dagger}a_{i}a_{i}b_{j}
\to&
 2( \langle a_{i}^{\dagger}a_{i} \rangle a_{i}b_{j}
   +\langle a_{i}b_{j} \rangle a_{i}^{\dagger}a_{i}
   -\langle a_{i}^{\dagger}a_{i} \rangle \langle a_{i}b_{j} \rangle)
\\ \nonumber
 a_{i}^{\dagger}b_{j}^{\dagger}b_{j}^{\dagger}b_{j}
\to&
 2( \langle a_{i}^{\dagger}b_{j}^{\dagger} \rangle b_{j}^{\dagger}b_{j}
   +\langle b_{j}^{\dagger}b_{j} \rangle a_{i}^{\dagger}b_{j}^{\dagger}
   -\langle a_{i}^{\dagger}b_{j}^{\dagger} \rangle \langle b_{j}^{\dagger}b_{j} \rangle)
\\ \nonumber
 a_{i}^{\dagger}a_{i}^{\dagger}a_{i}b_{j}^{\dagger}
\to&
 2( \langle a_{i}^{\dagger}a_{i} \rangle a_{i}^{\dagger}b_{j}^{\dagger}
   +\langle a_{i}^{\dagger}b_{j}^{\dagger} \rangle a_{i}^{\dagger}a_{i}
   -\langle a_{i}^{\dagger}a_{i} \rangle \langle a_{i}^{\dagger}b_{j}^{\dagger} \rangle)
\\
 a_{i}b_{j}^{\dagger}b_{j}b_{j}
\to&
 2( \langle a_{i}b_{j} \rangle b_{j}^{\dagger}b_{j}
   +\langle b_{j}^{\dagger}b_{j} \rangle a_{i}b_{j}
   -\langle a_{i}b_{j} \rangle \langle b_{j}^{\dagger}b_{j} \rangle)
\end{align}
where $\langle \cdots \rangle$ denotes the quantum average in the magnon vacuum. 
Here, we have omitted normal order of the quartic terms 
and assumed that 
$
 \langle a_{i}^{\dagger}b_{j} \rangle
=\langle a_{i}b_{j}^{\dagger} \rangle
=\langle a_{i}a_{i} \rangle
=\langle a_{i}^{\dagger}a_{i}^{\dagger} \rangle
=\langle b_{j}^{\dagger}b_{j}^{\dagger} \rangle
=\langle b_{j}b_{j} \rangle
=0
$ due to the conservation of magnetization. 
After the decomposition of the quartic terms, we have a 
quadratic form spin-wave Hamiltonian in real space. 
Carrying out the Bogoliubov transformation, we can diagonalize 
the quadratic spin-wave Hamiltonian into 
\begin{align}
 H_{\rm SW}^{\prime}
=
 \sum_{k=1}^{n_{\alpha}}\varepsilon_{k}^{(\alpha)}\alpha_{k}^{\dagger}\alpha_{k}
+\sum_{l=1}^{n_{\beta}}\varepsilon_{l}^{(\beta)}\beta_{l}^{\dagger}\beta_{l}
+E_{\rm GS}
\end{align}
where $\varepsilon_{k}^{(\alpha)}$ [$\varepsilon_{l}^{(\beta)}$] is 
the eigenvalue of the bosonic quasiparticle mode $\alpha_{k}$ ($\beta_{l}$), 
$n_{\alpha}$ ($n_{\beta}$) is the number of the $\alpha_{k}$ ($\beta_{l}$) modes, 
and $E_{\rm GS}$ is the ground-state energy.

\section{Effective magnetic Raman operator\\}

The magnetic Raman scattering is described by interaction between spin and photon. 
In this section, we follow a microscopic description of the magnetic Raman scattering, 
which is first given by Shastry and Shraiman \cite{ShastryPRL,Shastry,Ko,Michaud}, 
and present effective magnetic Raman operator on the Penrose lattice. 
First, we consider a strongly correlated single-band Hubbard model: 
\begin{align}
 H_{\rm Hb}
=H_{U}+H_{t}
=
 U\sum_{i}n_{i\uparrow}n_{i\downarrow}
-\sum_{i,j,\sigma}t_{ij}c_{i\sigma}^{\dagger}c_{j\sigma}
\end{align}
where $c_{i\sigma}^{\dagger}$ ($c_{i\sigma}$) is the electron creation (annihilation) 
operator at site $i$ with spin $\sigma=\uparrow, \downarrow$ 
and $n_{i\sigma}\equiv c_{i\sigma}^{\dagger}c_{i\sigma}$. 
$t_{ij}$ is the transfer integral, 
and $U ( > 0)$ is the on-site Coulomb repulsion. 
Hereafter, we restrict that electron hopping only occurs between nearest-neighbor sites.

The electron-photon coupling can be introduced by the Peierls substitution: 
$
 c_{i\sigma}^{\dagger}c_{j\sigma}
\to
 c_{i\sigma}^{\dagger}c_{j\sigma}
 \exp(\frac{ie}{\hbar c}\int_{j}^{i}{\bm A}\cdot d{\bm r})
$, 
where ${\bm A}$ is the photon vector potential. 
We assume that incoming and outgoing photon wavelengths are much larger than 
lattice spacing. 
Then second-quantized vector potential is written as 
$
 {\bm A}
= g_{\rm in}{\bm e}_{\rm in}\gamma_{{\bm k}_{\rm in}}
 +g_{\rm sc}{\bm e}_{\rm sc}^{*}\gamma_{{\bm k}_{\rm sc}}^{\dagger}
$
where 
$g_{\rm in}=\sqrt{hc^{2}/\omega_{\rm in}V}$ and 
$g_{\rm sc}=\sqrt{hc^{2}/\omega_{\rm sc}V}$ with volume $V$. 
$\omega_{\rm in} (\omega_{\rm sc})$, 
${\bm k}_{\rm in} ({\bm k}_{\rm sc})$, and 
${\bm e}_{\rm in} ({\bm e}_{\rm sc})$ stand for 
frequency, momentum, and polarization of incident (scattered) photon, respectively. 
$\gamma^{\dagger} (\gamma)$ 
denotes the photon creation (annihilation) operator. 
Expanding the exponential of the hopping terms, the current operator reads 
\begin{align}
 H_{\rm c}
=-\frac{ie}{\hbar c}\sum_{i,j,\sigma}t_{ij}
 {\bm A}\cdot{\bm \delta}_{ij} c_{i\sigma}^{\dagger}c_{j\sigma}
\end{align}
where ${\bm \delta_{ij}}$ is the vector connecting sites $i$ and $j$. 

Since the Raman process is made of two photons (one photon in, one photon out),  
we consider second-order terms in ${\bm A}$. 
We are interested in half-filled ($\sum_{\sigma}\langle n_{i\sigma} \rangle=1$) and 
localized ($U \gg t$) system, 
$H_{\rm c}$ and $H_{t}$ can be treated as a perturbation. 
In this situation, 
initial states and final states belong to the ground-state manifold of singly occupied states. 
The effective Raman operator reads 
\begin{align}
\nonumber
 {\cal R}
=&{\cal P}H_{\rm c}\frac{1}{\varepsilon_{i}-H_{U}-H_{t}}H_{\rm c}{\cal P}
\\
=&
 {\cal P}H_{\rm c}\frac{1}{\varepsilon_{i}-H_{U}}
 \sum_{n=0}^{\infty}\left(H_{t}\frac{1}{\varepsilon_{i}-H_{U}}\right)^{n}
 H_{\rm c}{\cal P}
\label{effectiveRam1}
\end{align}
where $\varepsilon_{i}$ is the initial-state energy and 
${\cal P}$ is a projection operator to the spin-$1/2$ sector. 
Because of the electron-hole symmetry in the half-filled band, 
any term of odd $n$ vanishes in Eq. \eqref{effectiveRam1}. 
Finally, we convert to electron operators into $S=1/2$ spin operators using 
the following projection: 
\begin{align}
 {\cal P}c_{i\sigma}^{\dagger}c_{i\sigma'}{\cal P}
=
 \frac{1}{2}\delta_{\sigma',\sigma}
+{\bm S}_{i}\cdot{\bm \tau}_{\sigma'\sigma}
\end{align}
where ${\bm \tau}$ is the Pauli matrix.

The second-order perturbation is the lowest nonvanishing order in the 
Shastry-Shraiman formulation, and it gives 
the Loudon-Fleury magnetic Raman operator \cite{Fleury}: 
\begin{align}
 {\cal R}^{(2)}
=
 \sum_{\langle i,j \rangle}
 \frac{4t^{2}}{U-\hbar\omega_{\rm in}}
 ({\bm e}_{\rm in}\cdot{\bm \delta}_{ij})
 ({\bm e}_{\rm sc}^{*}\cdot{\bm \delta}_{ij})
 {\bm S}_{i}\cdot{\bm S}_{j}
\label{LFoperator}
\end{align}
Here, we omit some constants, which does not affect the Raman intensity.

The fourth-order effective magnetic Raman operator, 
which is the next nonvanishing perturbation 
at the prefactor $t^{4}/(U-\hbar\omega_{\rm in})^{3}$, 
includes the scalar-spin-chirality terms 
${\bm S}_{i}\cdot({\bm S}_{j}\times{\bm S}_{k})$ 
and/or the ring-exchange terms 
$({\bm S}_{i}\cdot{\bm S}_{j})({\bm S}_{k}\cdot{\bm S}_{l})$. 
For details about the fourth-order magnetic Raman operator, 
see Appendix. 
If the incident photon energy $\hbar\omega_{\rm in}$ approaches to 
resonant limit $|U-\hbar\omega_{\rm in}|\sim t$, 
higher-order contributions can manifest in Raman intensities. 
%

For theoretical calculations, it is convenient to decompose the polarization dependence 
of the magnetic Raman spectrum into the irreducible representations (irreps) of the lattice point group. 
The point group of the Penrose lattice is ${\bf C}_{5v}$, 
polarization dependence of Raman active modes decomposes into 
two one-dimensional irreps $A_{1}$ and $A_{2}$, 
and one two-dimensional irrep $E_{2}$ as follow: 
\begin{align}
\nonumber
 A_{1}: e_{\rm in}^{x}e_{\rm sc}^{*x}+e_{\rm in}^{y}e_{\rm sc}^{*y}
\\ \nonumber
 A_{2}: e_{\rm in}^{x}e_{\rm sc}^{*y}-e_{\rm in}^{y}e_{\rm sc}^{*x}
\\ \nonumber
 E_{2}^{(1)}: e_{\rm in}^{x}e_{\rm sc}^{*x}-e_{\rm in}^{y}e_{\rm sc}^{*y}
\\
 E_{2}^{(2)}: e_{\rm in}^{x}e_{\rm sc}^{*y}+e_{\rm in}^{y}e_{\rm sc}^{*x}
\end{align}

The magnetic Raman spectrum is given by Fermi's golden rule:
\begin{align}
 I(\omega)
=
 \sum_{n}\Bigl|\langle \Psi_{n}|{\cal R}|\Psi_{0} \rangle\Bigr|^{2}
 \delta(\hbar\omega-E_{n}+E_{0})
\end{align}
where $|\Psi_{0}\rangle$ is a ground state of the Heisenberg model, 
$|\Psi_{n}\rangle$ is excited states, 
$E_{0}$ and $E_{n}$ are eigenvalues of ground state and excited states, respectively.

\section{Results}

\subsection{Second-order magnetic Raman intensity: Within the Loudon-Fleury mechanism\\}

First, we consider the Raman spectrum within the Loudon-Fleury mechanism. 
In this section, 
we use the second-order magnetic Raman operator [Eq. \eqref{LFoperator}]. 
Spin operators in Eq. \eqref{LFoperator} are expanded by the Holstein-Primakoff bosons. 
In this study, we consider the two-magnon scattering 
which corresponds to the expansion of the magnetic Raman operator 
up to the bosonic two-body terms. 

\begin{figure}[htb]
\centering
\includegraphics*[width=.7\linewidth]{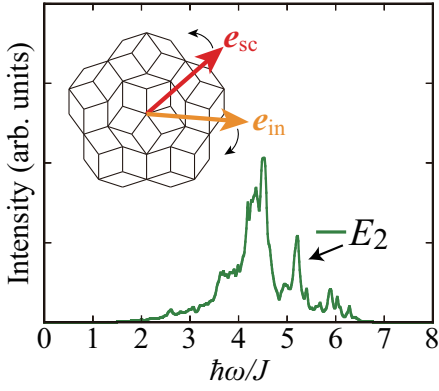}
\caption{
 Two-magnon scattering magnetic Raman spectrum of the $N=601$ sites cluster 
 Penrose lattice Heisenberg antiferromagnet 
 within the second-order magnetic Raman operator. 
 The spectrum comes from the $E_{2}$ representation of the ${\bf C}_{5v}$ point group, 
 and does not depend on the incident and scattered photon polarizations ${\bm e}_{\rm in}$ and ${\bm e}_{\rm sc}^{*}$. 
}
\label{Penrose2ndLF}
\end{figure}

In Fig. \ref{Penrose2ndLF}, 
we give the result of the spin-wave calculation of the
two-magnon scattering magnetic Raman intensity within the second-order Raman operator 
for the $N=601$ sites cluster of the Penrose lattice. 
We find the second-order Raman intensity comes from the $E_{2}$ representation 
and shows no linear polarization dependence. 
To understand this depolarization, 
we set the incident and scattered polarization vectors as
\begin{align}
 {\bm e}_{\rm in}
=(\cos\theta_{\rm in},\sin\theta_{\rm in}),
\;\;\;
 {\bm e}_{\rm sc}
=(\cos\theta_{\rm sc},\sin\theta_{\rm sc})
\end{align}
where $\theta_{\rm in}$ and $\theta_{\rm sc}$ are the angles of the polarization vectors 
of the incident and scattered photons with respect to the $x$ axis. 
Under this condition, 
the $E_{2}$ mode Raman spectrum is written as 
\begin{align}
\nonumber
 I(\omega,\theta_{\rm in},\theta_{\rm sc})
=&
 \sum_{n}\Bigl|\langle \Psi_{n}|
   {\cal R}_{E_{2}^{(1)}}\cos(\theta_{\rm in}+\theta_{\rm sc})
\\ \nonumber
&
  +{\cal R}_{E_{2}^{(2)}}\sin(\theta_{\rm in}+\theta_{\rm sc})
 |\Psi_{0} \rangle\Bigr|^{2}
\\
&\times
 \delta(\hbar\omega-E_{n}+E_{0})
\end{align}
where ${\cal R}_{E_{2}^{(1)}}$ and ${\cal R}_{E_{2}^{(2)}}$ are 
irreducible decomposed Raman operators of 
first- and second components of the $E_{2}$ representations, respectively. 
${\cal R}_{E_{2}^{(1)}}$ and ${\cal R}_{E_{2}^{(2)}}$ are 
orthogonal to each other, and 
cross sections of ${\cal R}_{E_{2}^{(1)}}$ and ${\cal R}_{E_{2}^{(2)}}$ are
degenerate. 
Therefore, the $E_{2}$ mode Raman spectrum is invariant to polarization angles: 
\begin{align}
\nonumber
 I(\omega,\theta_{\rm in},\theta_{\rm sc})
=&
 \cos^{2}(\theta_{\rm in}+\theta_{\rm sc})I_{E_{2}}(\omega)
\\
&
+\sin^{2}(\theta_{\rm in}+\theta_{\rm sc})I_{E_{2}}(\omega)
=
 I_{E_{2}}(\omega)
\end{align}
where $I_{E_{2}}(\omega)$ denotes intensity of the $E_{2}$ mode.

\subsection{Fourth-order magnetic Raman intensity: Beyond the Loudon-Fleury mechanism\\}

\begin{figure}[htb]
\centering
\includegraphics*[width=.7\linewidth]{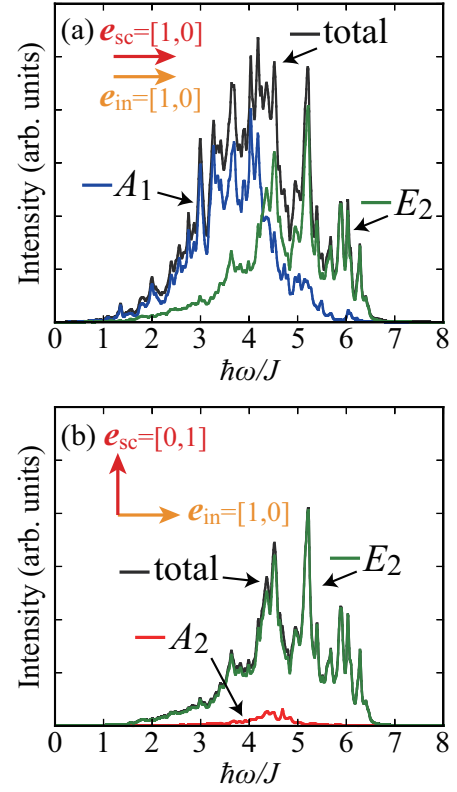}
\caption{
 Two-magnon scattering magnetic Raman spectra 
 of the fourth-order magnetic Raman operator 
 on the $N=601$ sites cluster for $xx$ polarization (a) and $xy$ polarization (b). 
 Sum of all Raman active modes (denoted by ``total'') is observed in each polarization 
 in actuality. 
}
\label{Penrose4thBLF}
\end{figure}

Next, we calculate the two-magnon Raman intensity 
of the fourth-order magnetic Raman operator. 
We consider two polarizations, 
one is called $xx$ polarization that corresponds to 
$(\theta_{\rm in},\theta_{\rm sc})=(0,0)$ [Fig. \ref{Penrose4thBLF}(a)], and 
another is called $xy$ polarization that corresponds to 
$(\theta_{\rm in},\theta_{\rm sc})=(0,\pi/2)$ [Fig. \ref{Penrose4thBLF}(b)]. 
As shown in Fig. \ref{Penrose4thBLF}, 
the fourth-order Raman operators yield spectral weight of 
the $A_{1}$ mode in the $xx$ polarization, and 
the $A_{2}$ mode in the $xy$ polarization, 
as well as the linearly polarization independent $E_{2}$ mode. 
The observed spectra of each polarization are written as 
$I_{xx}(\omega)=I_{A_{1}}(\omega)+I_{E_{2}}(\omega)$ for the $xx$ polarization, and 
$I_{xy}(\omega)=I_{A_{2}}(\omega)+I_{E_{2}}(\omega)$ for the $xy$ polarization. 
In general, the linear polarization dependence of the fourth-order Raman intensity 
is given by
\begin{align}
\nonumber
 I(\omega,\theta_{\rm in},\theta_{\rm sc})
=&
 \cos^{2}(\theta_{\rm in}-\theta_{\rm sc})I_{A_{1}}(\omega)
\\
&
+\sin^{2}(\theta_{\rm in}-\theta_{\rm sc})I_{A_{2}}(\omega)
+I_{E_{2}}(\omega)
\label{polardep4th}
\end{align}
As shown in Eq. \eqref{polardep4th}, 
the fourth-order Raman spectrum is observed as combination of irreducible spectra. 
To extract every irreducible representation from observations, 
we employ two linearly and one circularly polarized lights. 
If we only consider linearly polarized lights, 
we lack degrees of freedom to separate every irreducible representation, 
so that why circularly polarized light is required. 
Solving the relations of the polarizations, 
we can separate every irreducible representation as: 
\begin{align}
\nonumber
 I_{A_{1}}(\omega)&=I_{xx}(\omega)-\frac{1}{2}I_{LR}(\omega)
\\ \nonumber
 I_{A_{2}}(\omega)&=I_{xy}(\omega)-\frac{1}{2}I_{LR}(\omega)
\\
 I_{E_{2}}(\omega)&=\frac{1}{2}I_{LR}(\omega)
\end{align}
where $I_{LR}(\omega)$ is the $LR$ polarization Raman intensity, 
which $LR$ signifies left circularly polarized incident photon 
${\bm e}_{\rm in}=\frac{1}{\sqrt{2}}(1,i)$ and 
right circularly polarized scattered photon ${\bm e}_{\rm sc}=\frac{1}{\sqrt{2}}(1,-i)$. 

The fourth-order magnetic Raman spectra include 
contributions of several magnetic excitations. 
In particular, the $A_{2}$ mode spectrum is intriguing, 
because it provides a direct observation of dynamical spin-chirality fluctuations. 
The spin-chirality terms in the fourth-order magnetic Raman operator cancel 
on the two types of fourth-order-electron-hopping pathways: 
(1) four-site loop pathway and (2) three-site straight pathway. 
For example, we consider the two-dimensional periodic lattice 
with single-site unit cell. 
This lattice has only two primitive lattice vectors, and 
it always satisfies the conditions of the spin-chirality-term cancellation. 
However, this is not the case for the Penrose lattice. 
Because of the quasiperiodicity, the Penrose lattice has additional 
primitive lattice vectors and extra pathways of the fourth-order electron hopping, 
so that the spin-chirality-driven $A_{2}$ mode spectrum can survive.

\subsection{Effects of magnon-magnon interactions\\}

\begin{figure}[htb]
\centering
\includegraphics*[width=.7\linewidth]{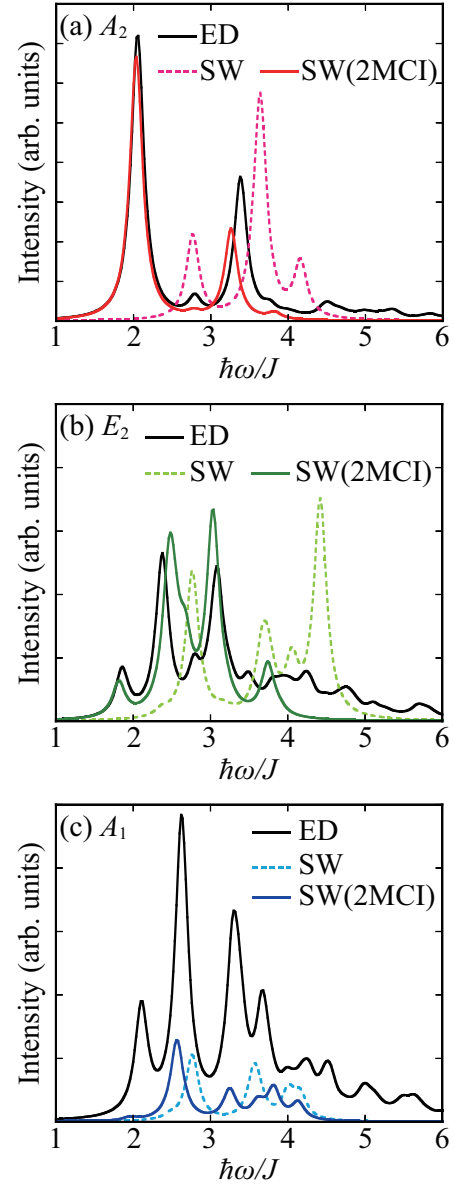}
\caption{
 The fourth-order magnetic Raman spectra on the $N=16$ sites cluster 
 for (a) $A_{2}$ mode, (b) $E_{2}$ mode, and (c) $A_{1}$ mode. 
 Spectra are calculated by Lancz\"{o}s exact diagonalization (ED), 
 spin-wave theory without magnon-magnon interactions (SW), and 
 spin-wave theory with magnon-magnon interactions introduced by 
 two-magnon excitation CI method [SW(2MCI)]. 
}
\label{N16magnoninteraction}
\end{figure}

In this section, we discuss effects of magnon-magnon interactions 
on the magnetic Raman scattering. 
We consider small size ($N=16$) cluster in order to compare 
spin-wave results with exact spectra obtained by the Lancz\"{o}s exact diagonalization. 
In the Lancz\"{o}s method, the Raman spectrum is obtained from 
a continued fraction: 
\begin{align}
 I(\omega)
=
-\frac{1}{\pi}{\rm Im}\left\{
 \langle \Psi_{0}|{\cal R}^{\dagger}
  \frac{1}{\hbar\omega+E_{0}+i\eta-H}
 {\cal R}|\Psi_{0}\rangle\right\}
\end{align}
where $\eta$ is a small imaginary part added to give a finite damping of the 
$\delta$-functions.

In the spin-wave calculation, we introduce the magnon-magnon interactions by 
the configuration interaction (CI) method. 
We apply the two-magnon excitation CI method in this study. 
We consider a zero-magnon state $|{\rm 0M}\rangle$ and 
two-magnon excited states $|{\rm 2M}\rangle$: 
\begin{align}
 |{\rm 0M}\rangle=|0\rangle,
\;\;\;
 |{\rm 2M}(k,l)\rangle=\alpha_{k}^{\dagger}\beta_{l}^{\dagger}|0\rangle
\end{align}
where $|0\rangle$ is a magnon-vacuum state. 
Spin-wave eigenstates are improved as 
\begin{align}
 |\Psi_{n}\rangle_{\rm CI}
=
 c_{0,n}|0\rangle +\sum_{k,l}c_{(k,l),n}\alpha_{k}^{\dagger}\beta_{l}^{\dagger}|0\rangle
\end{align}
Coefficients $c_{0,n}$ and $c_{(k,l),n}$ are obtained by diagonalization of 
the CI Hamiltonian matrix: 
\begin{align}
 H_{\rm CI}
=\left[
 \begin{array}{cc}
  \langle{\rm 0M}|H_{\rm SW}|{\rm 0M}\rangle & \langle{\rm 0M}|H_{\rm SW}|{\rm 2M}\rangle
 \\
  \langle{\rm 2M}|H_{\rm SW}|{\rm 0M}\rangle & \langle{\rm 2M}|H_{\rm SW}|{\rm 2M}\rangle
 \end{array}
 \right]
\end{align}
We note that the two-magnon excitation CI calculation 
corresponds to solving the ladder-approximation Bethe-Salpeter equation 
with interactions of the quartic magnon terms \cite{Canali}.

Results are shown in Fig. \ref{N16magnoninteraction}. 
First, we focus on the $A_{2}$ mode spectra [Fig. \ref{N16magnoninteraction}(a)]. 
Comparing the both spin-wave results, 
Raman peaks soften down after considering the magnon-magnon interactions. 
The line shape and peak positions of the result of the interacting spin-wave 
are in good agreement with result of the exact diagonalization. 
We conclude that the spin-wave calculation of the two-magnon scattering process 
can describe the spin-chirality-driven $A_{2}$ mode magnetic Raman spectrum very well. 

For the $E_{2}$ mode spectra [Fig. \ref{N16magnoninteraction}(b)], 
the two-magnon scattering intensity of the spin-wave theory 
with the magnon-magnon interactions agrees with the exact result 
at low-frequency (about $\hbar\omega < 4J$) part . 
However, the interacting spin-wave result lacks 
high-frequency tail of the exact result, which 
is expected higher-order contributions. 

On the other hand, from Fig. \ref{N16magnoninteraction}(c), 
the $A_{1}$ mode magnetic Raman spectra 
of the exact diagonalization and spin-waves disagree.
The two-magnon scattering spin-wave spectra are quite smaller than the exact spectrum, 
even if it includes the magnon-magnon interactions. 
This suggests that the higher order multimagnon scattering, 
for instance four-magnon scattering, 
is dominant in the $A_{1}$ mode Raman intensity.

\section{Conclusion\\}

We have presented the magnetic Raman spectra of the two-dimensional 
${\bf C}_{5v}$ Penrose lattice Heisenberg antiferromagnets. 
The Raman intensity within the Loudon-Fleury mechanism comes from 
the $E_{2}$ representation and shows no linear polarization dependence 
due to the degeneracy of the two-dimensional irreducible representation $E_{2}$. 
In contrast, the fourth-order Raman operator yields spectral weights of 
$A_{1}$ and $A_{2}$, as well as $E_{2}$, representations and 
therefore exhibit strong polarization dependence 
in the Raman intensities beyond the Loudon-Fleury mechanism. 
The $A_{2}$ mode spectrum is driven by scalar-spin-chirality terms, and 
it is arisen from quasiperiodic structure of the Penrose lattice. 
We can separately extract every irreducible representation 
from the observation with the use of two linearly and one circularly polarized lights. 
The two-magnon scattering with the magnon-magnon interactions can describe 
the $A_{2}$ and $E_{2}$ mode spectra very well. 
This means that the spin-chirality excitations and exchange excitations 
can be mainly understood by the two-magnon scattering process. 
In contrast, the $A_{1}$ mode spectrum, which is almost caused by 
the ring-exchange excitations, disagree with the two-magnon scattering result. 
To understand the $A_{1}$ mode Raman spectrum, 
we have to consider the multimagnon scattering process, 
which is left for further investigation.

\begin{acknowledgement}
We would like to thank J. Ohara and Y. Noriki for useful discussions. 
This study was supported by the Ministry of Education, Culture, Sports, 
Science, and Technology of Japan. 
\end{acknowledgement}

\section*{Appendix\\}

In this section, we shall present the details of 
the fourth-order magnetic Raman operator. 
It is obtained by a fourth-order perturbation: 
\begin{align*}
 {\cal R}^{(4)}
={\cal P}H_{c}\frac{1}{\varepsilon_{i}-H_{U}}
 H_{t}\frac{1}{\varepsilon_{i}-H_{U}}
 H_{t}\frac{1}{\varepsilon_{i}-H_{U}}H_{c}{\cal P}
\end{align*}
where $H_{c}$ is the current operator, $H_{t}$ is the electron transfer operator, 
and $H_{U}$ is the on-site Coulomb repulsion operator, respectively. 
$\varepsilon_{i}$ is the energy of the initial state. 
We fix that the initial states are 
direct product of singly-occupied electron states with 
incident photon, and the intermediate states are 
one holon and one doublon states with no photons. 
Under this condition, 
$(\varepsilon_{i}-H_{U})^{-1}=(\hbar\omega_{\rm in}-U)^{-1}$ becomes 
a $c$-number. 
%
${\cal P}$ is the projection operator which converts electron operators 
into spin-$1/2$ operators. 

\begin{figure}[htb]
\centering
\includegraphics*[width=.67\linewidth]{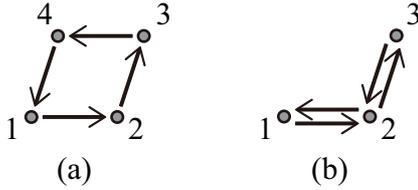}
\caption{
 Two types of fourth-order-electron-hopping pathways. 
 (a) Four-site loop pathway and (b) three-site pathway. 
 Arrows indicate the movement of electrons arising from $H_{c}$ and $H_{t}$. 
}
\label{path4th}
\end{figure}

The fourth-order effective magnetic Raman operator is written as 
\begin{align*}
 {\cal R}^{(4)}
=&
 \sum_{\langle 1,2,3,4 \rangle}
 \frac{t^{4}}{(U-\hbar\omega_{\rm in})^{3}}\biggl\{
\\
&
  -4\sum_{n=1}^{4}({\bm e}_{\rm in}\cdot{\bm \delta}_{n})
   ({\bm e}_{\rm sc}^{*}\cdot[{\bm \delta}_{n+1}+2{\bm \delta}_{n+2}+{\bm \delta}_{n+3}])
\\
&\times
  \Bigl[{\cal Q}_{1234}+{\cal Q}_{1432}-{\cal Q}_{1324}\Bigr]
\\
&+2i\sum_{n=1}^{4}\Delta_{n}^{\rm ch}
 {\bm S}_{n+2}\cdot({\bm S}_{n+1}\times{\bm S}_{n})
\\
&+\sum_{n=1}^{4}\Delta_{n}^{\rm ex}{\bm S}_{n}\cdot{\bm S}_{n+1}
 +\sum_{n=1}^{2}\Delta_{n}^{\rm ex^{\prime}}{\bm S}_{n}\cdot{\bm S}_{n+2}
\biggr\}
\\
&+\sum_{\langle 1,2,3 \rangle}
  \frac{t^{4}}{(U-\hbar\omega_{\rm in})^{3}}\biggl\{
\\
&
 4i\Bigl[({\bm e}_{\rm in}\cdot{\bm \delta}_{1})
         ({\bm e}_{\rm sc}^{*}\cdot{\bm \delta}_{2})
        -({\bm e}_{\rm in}\cdot{\bm \delta}_{2})
         ({\bm e}_{\rm sc}^{*}\cdot{\bm \delta}_{1})\Bigr]
\\
&\times
 {\bm S}_{3}\cdot({\bm S}_{2}\times{\bm S}_{1})
\\
&+2\sum_{n=1}^{2}\tilde{\Delta}_{n}^{\rm ex}
 {\bm S}_{n}\cdot{\bm S}_{n+1}
\\
&
-2\Bigl[({\bm e}_{\rm in}\cdot{\bm \delta}_{1})
        ({\bm e}_{\rm sc}^{*}\cdot{\bm \delta}_{2})
       +({\bm e}_{\rm in}\cdot{\bm \delta}_{2})
        ({\bm e}_{\rm sc}^{*}\cdot{\bm \delta}_{1})\Bigr]
\\
&\times
 {\bm S}_{1}\cdot{\bm S}_{3}
\biggr\}
\end{align*}
\begin{align*}
 {\cal Q}_{ijkl}
\equiv
 ({\bm S}_{i}\cdot{\bm S}_{j})({\bm S}_{k}\cdot{\bm S}_{l})
\end{align*}
\begin{align*}
 \Delta_{n}^{\rm ch}
\equiv&
  ({\bm e}_{\rm in}\cdot{\bm \delta}_{n})
  ({\bm e}_{\rm sc}^{*}\cdot[-{\bm \delta}_{n+1}-2{\bm \delta}_{n+2}+{\bm \delta}_{n+3}])
\\
&
 +
  ({\bm e}_{\rm in}\cdot{\bm \delta}_{n+1})
  ({\bm e}_{\rm sc}^{*}\cdot[-{\bm \delta}_{n+2}+2{\bm \delta}_{n+3}+{\bm \delta}_{n}])
\\
&+({\bm e}_{\rm in}\cdot{\bm \delta}_{n+2})
  ({\bm e}_{\rm sc}^{*}\cdot[{\bm \delta}_{n+3}+2{\bm \delta}_{n}+{\bm \delta}_{n+1}])
\\
&
 +
  ({\bm e}_{\rm in}\cdot{\bm \delta}_{n+3})
  ({\bm e}_{\rm sc}^{*}\cdot[-{\bm \delta}_{n}-2{\bm \delta}_{n+1}-{\bm \delta}_{n+2}])
\end{align*}
\begin{align*}
 \Delta_{n}^{\rm ex}
\equiv&
  ({\bm e}_{\rm in}\cdot{\bm \delta}_{n})
  ({\bm e}_{\rm sc}^{*}\cdot[-{\bm \delta}_{n+1}+2{\bm \delta}_{n+2}-{\bm \delta}_{n+3}])
\\
&
 +
  ({\bm e}_{\rm in}\cdot{\bm \delta}_{n+1})
  ({\bm e}_{\rm sc}^{*}\cdot[{\bm \delta}_{n+2}-2{\bm \delta}_{n+3}-{\bm \delta}_{n}])
\\
&+({\bm e}_{\rm in}\cdot{\bm \delta}_{n+2})
  ({\bm e}_{\rm sc}^{*}\cdot[{\bm \delta}_{n+3}+2{\bm \delta}_{n}+{\bm \delta}_{n+1}])
\\
& +
  ({\bm e}_{\rm in}\cdot{\bm \delta}_{n+3})
  ({\bm e}_{\rm sc}^{*}\cdot[-{\bm \delta}_{n}-2{\bm \delta}_{n+1}+{\bm \delta}_{n+2}])
\end{align*}
\begin{align*}
\Delta_{n}^{\rm ex^{\prime}}
\equiv&
  ({\bm e}_{\rm in}\cdot{\bm \delta}_{n})
  ({\bm e}_{\rm sc}^{*}\cdot[{\bm \delta}_{n+1}+2{\bm \delta}_{n+2}-{\bm \delta}_{n+3}])
\\
& +
  ({\bm e}_{\rm in}\cdot{\bm \delta}_{n+1})
  ({\bm e}_{\rm sc}^{*}\cdot[-{\bm \delta}_{n+2}+2{\bm \delta}_{n+3}+{\bm \delta}_{n}])
\\
&+({\bm e}_{\rm in}\cdot{\bm \delta}_{n+2})
  ({\bm e}_{\rm sc}^{*}\cdot[{\bm \delta}_{n+3}+2{\bm \delta}_{n}-{\bm \delta}_{n+1}])
\\
& +
  ({\bm e}_{\rm in}\cdot{\bm \delta}_{n+3})
  ({\bm e}_{\rm sc}\cdot[-{\bm \delta}_{n}+2{\bm \delta}_{n+1}+{\bm \delta}_{n+2}])
\end{align*}
\begin{align*}
\tilde{\Delta}_{n}^{\rm ex}
\equiv&
   ({\bm e}_{\rm in}\cdot{\bm \delta}_{n})
   ({\bm e}_{\rm sc}^{*}\cdot[{\bm \delta}_{1}+{\bm \delta}_{2}])
\\
&
  +({\bm e}_{\rm in}\cdot[{\bm \delta}_{1}+{\bm \delta}_{2}])
   ({\bm e}_{\rm sc}^{*}\cdot{\bm \delta}_{n})
\end{align*}
where $\sum_{\langle 1,2,3,4 \rangle}$ is taken over four-sites loop pathways, and 
$\sum_{\langle 1,2,3 \rangle}$ is taken over three-sites linked pathways 
(see Fig. \ref{path4th}). 
${\bm e}_{\rm in}$ and ${\bm e}_{\rm sc}^{*}$ are the polarization vectors of
incident and scattered photons. 
${\bm \delta}_{n}\equiv{\bm r}_{n+1}-{\bm r}_{n}$ is the vector that connects 
site $n$ to site $n+1$. 
In these equations, we set $n\doteq n+4$ as a modulus.

%
%

\end{document}